\documentclass[twocolumn]{aastex62}

\usepackage[shortcuts]{extdash}

\def\apjl{ApJL}

\def\aap{A\&A}

\def\apjs{ApJS}

\def\swift{\textit{Swift}}
\def\chandra{\textit{Chandra}}
\def\xmm{\textit{XMM-Newton}}
\def\nustar{\textit{NuSTAR}}
\def\panstarrs{PanSTARRS}

\received{}
\revised{}
\accepted{}
\submitjournal{ApJ}

\shorttitle{The (re)appearance of NGC 925 ULX-3}
\shortauthors{Earnshaw et al.}

\begin{document}

\title{The (re)appearance of NGC 925 ULX-3, a new transient ULX}

\correspondingauthor{Hannah Earnshaw}
\email{hpearn@caltech.edu}

\author[0000-0001-5857-5622]{Hannah P. Earnshaw}
\affil{Cahill Center for Astronomy and Astrophysics, California Institute of Technology, Pasadena, CA 91125, USA}

\author{Marianne Heida}
\affil{European Southern Observatory, Karl-Schwarzschild-Strasse 2, 85748, Garching bei M\"{u}nchen, Germany}
\affil{Cahill Center for Astronomy and Astrophysics, California Institute of Technology, Pasadena, CA 91125, USA}

\author{Murray Brightman}
\affil{Cahill Center for Astronomy and Astrophysics, California Institute of Technology, Pasadena, CA 91125, USA}

\author{Felix F\"{u}rst}
\affil{Quasar Ltd for ESA, European Space Astronomy Centre (ESAC), Science Operations Department, E-28692, Villanueva de la Cañada, Madrid, Spain}

\author{Fiona A. Harrison}
\affil{Cahill Center for Astronomy and Astrophysics, California Institute of Technology, Pasadena, CA 91125, USA}

\author{Amruta Jaodand}
\affil{Cahill Center for Astronomy and Astrophysics, California Institute of Technology, Pasadena, CA 91125, USA}

\author{Matthew J. Middleton}
\affil{Department of Physics and Astronomy, University of Southampton, Highfield, Southampton SO17 1BJ, UK}

\author{Timothy P. Roberts}
\affil{Department of Physics, Centre for Extragalactic Astronomy, Durham University, South Road, Durham DH1 3LE, UK}

\author{Rajath Sathyaprakash}
\affil{Department of Physics, Centre for Extragalactic Astronomy, Durham University, South Road, Durham DH1 3LE, UK}

\author{Daniel Stern}
\affil{Jet Propulsion Laboratory, California Institute of Technology, 4800 Oak Grove Drive, Pasadena, CA 91109, USA}

\author{Dominic J. Walton}
\affil{Institute of Astronomy, Madingley Road, Cambridge CB3 0HA, UK}

\begin{abstract}

We report the discovery of a third ULX in NGC~925 (ULX-3), detected in November 2017 by \chandra\ at a luminosity of $L_{\rm X} = (7.8\pm0.8)\times10^{39}$\,erg\,s$^{-1}$. Examination of archival data for NGC~925 reveals that ULX-3 was detected by \swift\ at a similarly high luminosity in 2011, as well as by \xmm\ in January 2017 at a much lower luminosity of $L_{\rm X} = (3.8\pm0.5)\times10^{38}$\,erg\,s$^{-1}$. With an additional \chandra\ non-detection in 2005, this object demonstrates a high dynamic range of flux of factor $\gtrsim26$. In its high-luminosity detections, ULX-3 exhibits a hard power-law spectrum with $\Gamma=1.6\pm0.1$, whereas the \xmm\ detection is slightly softer, with $\Gamma=1.8^{+0.2}_{-0.1}$ and also well-fitted with a broadened disc model. The long-term light curve is sparsely covered and could be consistent either with the propeller effect or with a large-amplitude superorbital period, both of which are seen in ULXs, in particular those with neutron star accretors. Further systematic monitoring of ULX-3 will allow us to determine the mechanism by which ULX-3 undergoes its extreme variability and to better understand the accretion processes of ULXs. \\

\end{abstract}

\section{Introduction} \label{sec:intro}

Ultraluminous X-ray sources (ULXs) are defined as extragalactic non-nuclear X-ray point sources with luminosities in excess of $10^{39}$\,erg\,s$^{-1}$, the majority of which are widely thought to be stellar-mass compact objects accreting at super-Eddington rates (for a recent review see \citealt{kaaret17}). One key piece of evidence that these sources are accreting in a regime different to the canonical sub-Eddington states was the detection of a high-energy turnover in their 5--10\,keV spectrum, first seen in high-quality \xmm\ data (e.g. \citealt{stobbart06,gladstone09}) and later confirmed by observations with \nustar, whose coverage above 10\,keV established the presence of a turnover and steep power-law drop-off in the spectra of most ULXs above $\sim$5\,keV (e.g. \citealt{bachetti13,mukherjee15,walton14,walton15b}). 

Explicit confirmation of the stellar-mass nature of at least a proportion of ULXs came from the detection of coherent pulsations in M82 X-2 \citep{bachetti14}, allowing the compact object to be unambiguously identified as a neutron star. Since then, the population of known neutron star ULXs has risen to at least 7, both through the detection of pulsations \citep{fuerst16,israel17a,israel17b,carpano18,rodriguezcastillo19,sathyaprakash19} and through the detection of probable cyclotron resonance scattering spectral features \citep{brightman18}. Their extreme luminosities imply accretion rates of hundreds of times the Eddington rate for neutron stars, and they tend to exhibit pulsations with period $\sim$1\,s, with sinusoidal pulse profiles suggesting that they are not highly beamed. The low count rates due to being extragalactic sources, the presence of spin-up and orbital modulation, and possible transience in pulsations make them very challenging to detect, however, and the proportion of ULXs that contain neutron star accretors as opposed to black holes is still an open question (e.g. \citealt{king16,middleton17}), especially since their broadband spectra are otherwise indistinguishable from the rest of the ULX population (e.g. \citealt{koliopanos17,pintore17,walton18}).

Neutron star ULXs do share a property of extreme long-term variability of various types, often over orders of magnitude in flux. One example of such variability is sudden and large drops in brightness, leading to an approximately bimodal distribution in flux. This may be due to the onset of the propeller regime (e.g. \citealt{tsygankov16a}), in which the magnetospheric radius of the magnetic field is larger than the corotation radius of the accretion disc, creating a centrifugal barrier to mass accretion and causing the flux to drop correspondingly \citep{illarionov75,stella86}. This may be a way of identifying candidate neutron star ULXs in the absence of detected pulsations (\citealt{earnshaw18}; Song et al. subm.). 

Several neutron star ULXs also exhibit superorbital periods on the order of tens of days, varying by up to factors of tens in flux, discovered through long-term monitoring using \swift\ (e.g. \citealt{walton16,fuerst18,brightman19}). Several mechanisms have been proposed for their cause, including precession of a warped disk obscuring the central source (e.g. \citealt{motch14}), or Lense-Thirring precession of the supercritical accretion flow itself (e.g. \citealt{middleton18, middleton19}). Monitoring at a regular cadence for long durations is important, as a high-amplitude superorbital period with limited sampling may result in the source appearing to have a bimodal flux distribution, which could then be interpreted as a propeller transition (e.g. M82~X-2; \citealt{tsygankov16a,brightman19}). It is also possible for ULXs to exhibit both of these effects -- for example, NGC~5907~ULX-1, which has a $\sim$78 day superorbital period and also underwent a greater drop in flux potentially due to the propeller effect in 2013 \citep{walton15a,walton16,fuerst17,israel17a}. It is worth noting that the presence of a superorbital period in itself is not evidence of a neutron star accretor, as they can also be seen in black hole systems, such as the well-studied Galactic source Cyg~X-1 (e.g. \citealt{rico08}).

The transient nature of ULXs and the sparse temporal X-ray coverage of most galaxies means that more ULXs are being discovered all the time, without necessarily being `new' sources. Some discoveries, having never been detected before, have previous upper limits establishing that large increases in flux must have taken place (e.g. \citealt{pintore18a,earnshaw19,wang19}), although limited coverage generally means that recent periods of high luminosity cannot be ruled out. Others have been detected at luminosities below $10^{39}$\,erg\,s$^{-1}$ before undergoing an increase in brightness that takes them into the ULX regime (e.g. \citealt{hu18}). 

This paper investigates the appearance of a previously unknown ULX during the latest \chandra\ observation in 2017 of NGC~925, a spiral galaxy at 9.56\,Mpc (from Cepheids distances recorded in the NED Distances Database\footnote{https://ned.ipac.caltech.edu/Library/Distances/}; \citealt{steer17}). The galaxy is known to contain two previously studied ULXs \citep{pintore18b}; we report on the discovery of a third ULX at position 02$^h$\,27$^m$\,20$^s$.18, +33$^\circ$\,34$^\prime$\,12.84$^{\prime\prime}$ (J2000) which we designate NGC~925~ULX-3 (henceforth ULX-3 in this paper). Upon investigation of archival data for this source, we found that it was previously detected by \xmm\ at a sub-ULX luminosity, and was also detected at ULX luminosities by \swift\ in 2011. In this paper we present the results of our analysis of the spectral and long-term timing properties of ULX-3, and discuss how it fits into the broader picture of extreme variability in ULXs.

\section{Data Reduction \& Analysis} \label{sec:data}

\begin{figure*}
	\vspace{4mm}
	\begin{center}
	\includegraphics[height=6cm]{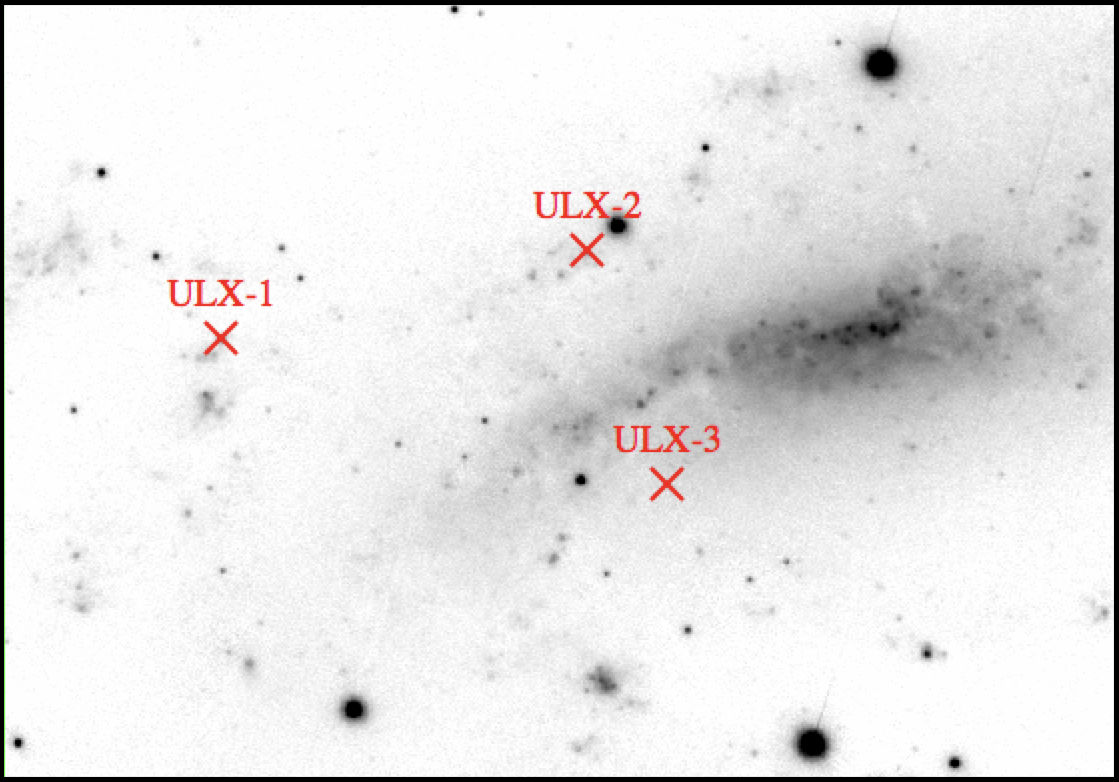}
	\hspace{3mm}
	\includegraphics[height=6cm]{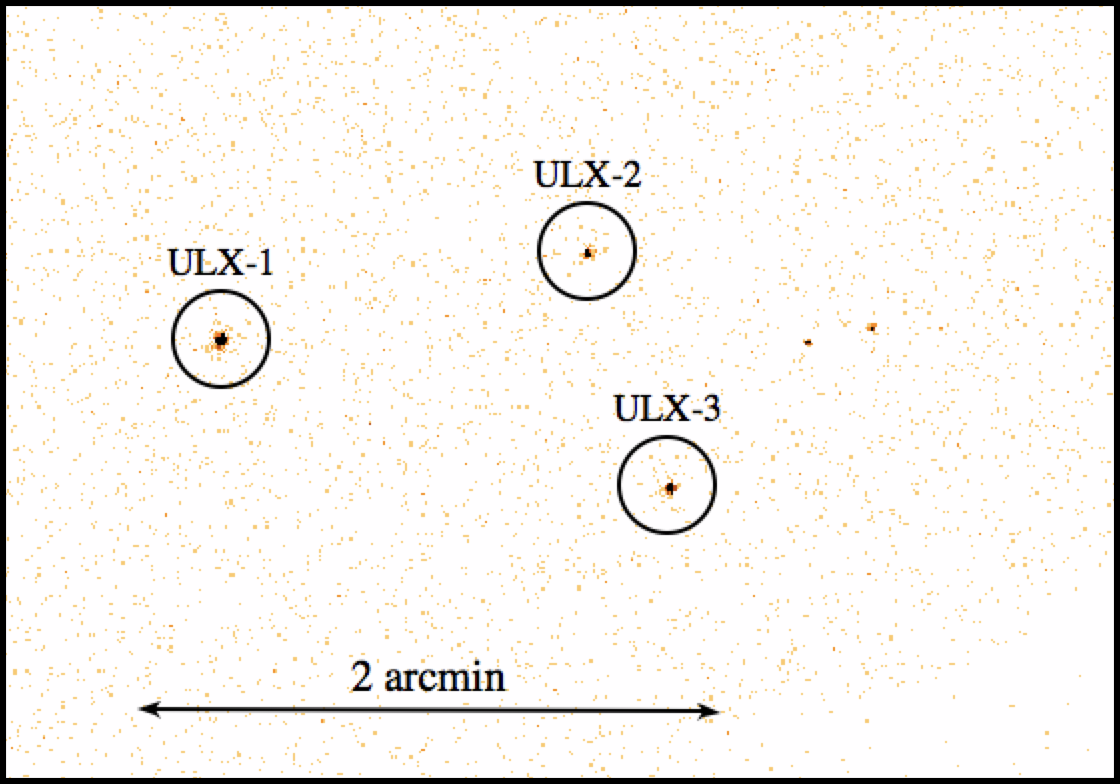}
	\end{center}
	\caption{\panstarrs\ $r$-band ({\it left}) and \chandra\ X-ray ({\it right}) images of spiral galaxy NGC~925, with the positions of three ULXs marked with red crosses and black circles. \label{fig:ngc925image}}
	\vspace{2mm}
\end{figure*}

\subsection{Chandra}

We observed NGC~925 using ACIS-S for 10\,ks on 2017 December 1, in order to obtain a precise localization of one of the other ULXs in the galaxy for comparison to near-infrared data (Heida et al. in prep). We reduced this observation (Obs. ID 20356; see Fig.~\ref{fig:ngc925image}, right) as well as an archival 2.2\,ks observation taken 2005 November 23 (Obs. ID 7104) using {\sc ciao} v4.10 with {\sc caldb} v4.7.9. We used the {\tt wavdetect} routine to determine the location of the new source, which we found to be 02$^h$\,27$^m$\,20$^s$.18, +33$^\circ$\,34$^\prime$\,12$^{\prime\prime}$.84 (J2000) with statistical 1$\sigma$ error of 0.03$^{\prime\prime}$. There were insufficient background sources for further astrometric correction to be performed. We subsequently used this position to extract data products from all other observations. Products were extracted from a circular region with radius 3$^{\prime\prime}$ centered on the source, with an annular background region centered on the source with inner radius 3$^{\prime\prime}$ and outer radius 20$^{\prime\prime}$. The spectrum and associated response and auxiliary files were extracted using the routine {\tt specextract}. The source was not detected in the archival observation, so we obtained a 3$\sigma$ upper limit on the flux using the {\tt srcflux} routine, assuming an absorbed power-law model with Galactic absorption ($N_{\rm H} = 7.26\times10^{20}$\,cm$^{-2}$; \citealt{willingale13}) and $\Gamma=2$. 

\subsection{XMM-Newton \& NuSTAR}

There is a single, 50\,ks archival \xmm\ observation of NGC~925 (Obs. ID 0784510301), taken on 2018 January 18. We extracted the data from the EPIC-pn and EPIC-MOS instruments using the \xmm\ {\sc sas} v17.0.0 software, producing calibrated event lists using the tasks {\tt emproc} and {\tt epproc}. Periods of high background flaring were removed by filtering out intervals of time during which the 10--12\,keV count rate exceeded 0.35\,cts/s across the EPIC-MOS detectors and 0.4\,cts/s across the EPIC-pn detector. We extracted data products from a 20$^{\prime\prime}$ radius circular source region, using a 40$^{\prime\prime}$ radius circular region on the same chip with a similar distance from the readout node for the background. Events with {\tt FLAG==0 \&\& PATTERN$<$4} were selected from EPIC-pn, and {\tt PATTERN$<$12} from EPIC-MOS. The tasks {\tt rmfgen} and {\tt arfgen} were used to create redistribution matrices and auxiliary response files respectively. 

The \xmm\ observation was taken quasi-simultaneously with a \nustar\ observation (Obs. ID 30201003002), though ULX-3 was not detected by \nustar\ at that time due to its low flux. We used XIMAGE to find a 3$\sigma$ upper limit on the count rate.

\subsection{Swift}

There are 18 archival observations of NGC~925 with \swift-XRT between 2011 and 2017 (Obs. ID: 00045596001--00045596018). Clean event lists were created using the FTOOLS task {\tt xrtpipeline} v0.13.4. Source and background spectra were extracted using the task {\tt xselect}, using a 30$^{\prime\prime}$ radius circular source region and a 70$^{\prime\prime}$ radius circular background region. Auxiliary response files were created using the task {\tt xrtmkarf} and the relevant redistribution matrix obtained from the CALDB. For those observations with sufficient numbers of counts to perform basic spectral fitting ($>$30), we obtain fluxes from the best-fitting absorbed power-law model using XSPEC v12.10.0 \citep{arnaud96}. Otherwise, we calculate fluxes from the count rate using PIMMS, assuming an absorbed power-law model with Galactic absorption and $\Gamma=1.6$ (consistent with the measured slope of the \chandra\ spectrum, see Section~\ref{sec:chandra}). 3$\sigma$ upper limits for \swift\ non-detections were determined using the {\tt sosta} routine in XIMAGE. 

\subsection{Optical/Infrared}

The deepest existing optical coverage of the part of NGC~925 that contains the ULXs is with PanSTARRS. The PanSTARRS catalogue \citep{flewelling16} contains one potential counterpart, a faint object only detected in the $r$ band with $m_r = 20.94\pm0.03$ mag at $0.4^{\prime\prime}$ from the \chandra\ position. There are no {\it WISE} or {\it Spitzer} sources coincident with the ULX. 

\begin{figure*}
	\begin{center}
	\includegraphics[width=16cm]{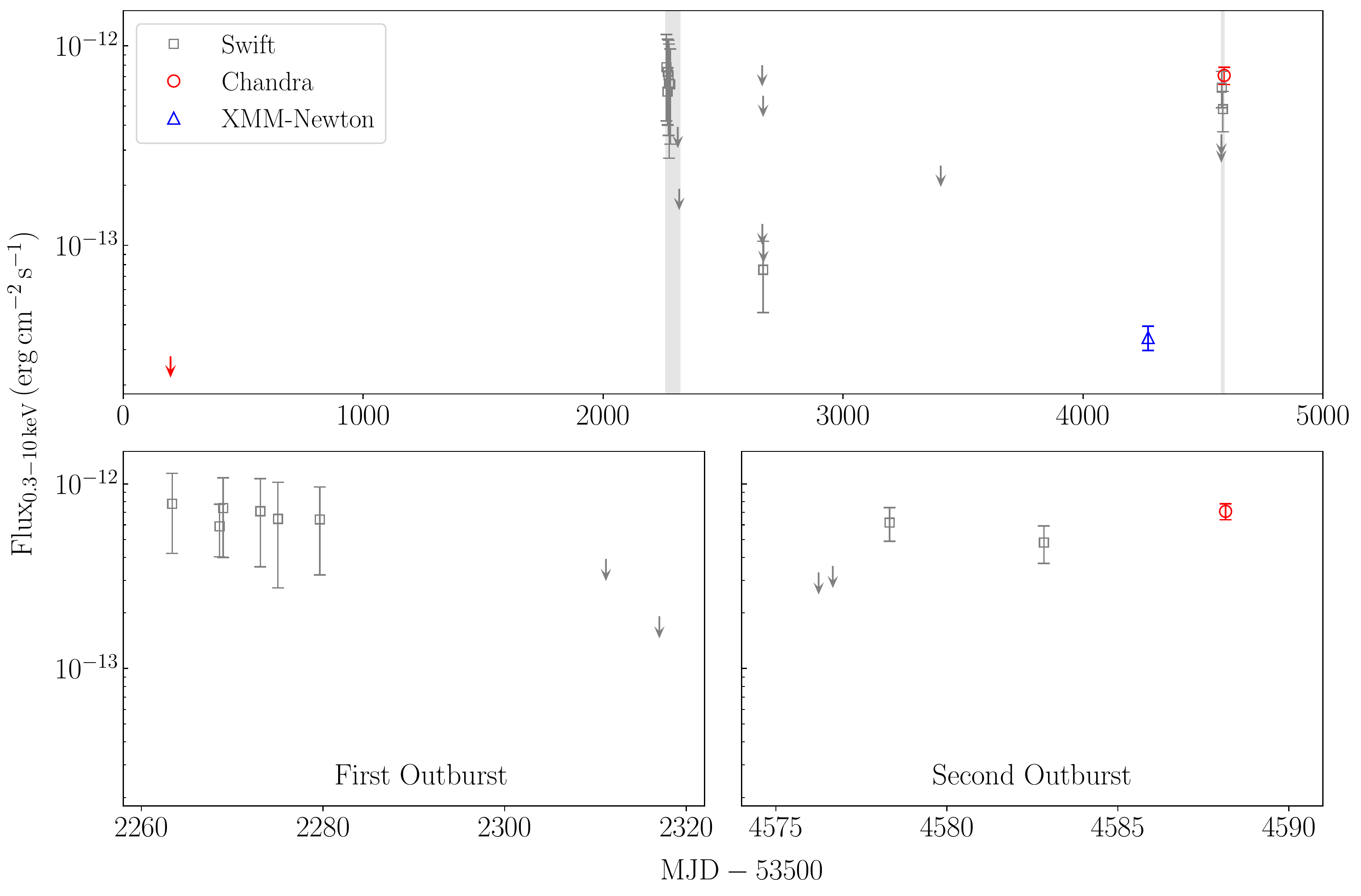}
	\end{center}
	\vspace{-5mm}
	\caption{Long term light curve of ULX-3, between the years 2005 and 2017. Detections are marked with empty symbols and 3$\sigma$ upper limits by downward arrows, with \chandra\ data in red (circles), \xmm\ in blue (triangle), and \swift\ in grey (squares). The first and second high flux epochs are shaded in grey and expanded in the lower subplots. \label{fig:longtermlc}}
\end{figure*}

\section{Results} \label{sec:results}

The long-term light curve of ULX-3 (Fig.~\ref{fig:longtermlc}), made up of \swift, \xmm, and \chandra\ data, shows that prior to the 2017 re-brightening in which we discovered the source, ULX-3 also underwent an outburst to super-Eddington luminosities in 2011, over the course of at least 17 days while the galaxy was being monitored by \swift. 

We fitted all spectra using XSPEC v12.10.0 \citep{arnaud96}, over the energy range 0.3--10\,keV. Given the generally low count rates, we grouped all spectra to a minimum of one count per bin and used background-subtracted Cash statistics \citep{cash79} when fitting. In all cases there is insufficient data to constrain any additional absorption beyond the Galactic value of $N_{\rm H} = 7.26\times10^{20}$\,cm$^{-2}$ \citep{willingale13} so we freeze the interstellar absorption to this value, fitting the absorption component using the {\tt tbabs} XSPEC model and \citet{wilms00} abundance tables throughout. Spectral fitting results are given in Table~\ref{tab:specfitresults}.

Since we are not using $\chi^2$ statistics, we instead use the Anderson-Darling test \citep{stephens74}, as implemented in XSPEC, to evaluate the goodness of fit for each best-fitting model. Upon running the {\tt goodness} command to perform Monte-Carlo simulations of the data based on the model parameters, the returned percentage is the proportion of simulations with a value of the test statistic lower than that for the data. Therefore, a value of $\lesssim$50\% indicates a model that fits the data well, and higher percentages indicate that the majority of simulations are better-fitted by the model than the data is (with a value of 95\% indicating that the null hypothesis of the data being described by the model can be rejected with a confidence of 2$\sigma$, and so on). All quoted errors are 90\% confidence intervals, calculated from the $\Delta C$ value when the parameter is varied.

\begin{deluxetable*}{@{}c@{}c@{~~}c@{}c@{~}c@{~}c@{~~}c@{}c@{~}c@{~}c@{}c@{}}
	\tablecaption{The spectral fitting results for X-ray observations of ULX\=/3. \label{tab:specfitresults}}
	\tablecolumns{11}
	\tablenum{1}
	\tablewidth{0pt}
	\tablehead{
		 \colhead{Dataset} & \colhead{Obs. Date} & \multicolumn{4}{c}{\tt tbabs * powerlaw} & \multicolumn{5}{c}{\tt tbabs * diskbb/diskpbb}  \\
		 \colhead{} & \colhead{} & \colhead{$\Gamma$} & \colhead{$F_{0.3-10\rm keV}$\tablenotemark{a}} & \colhead{$C/{\rm dof}$} & \colhead{A-D\tablenotemark{b}} & \colhead{$T_{\rm in}$ (keV)} & \colhead{$p$} & \colhead{$F_{0.3-10\rm keV}$\tablenotemark{a}} & \colhead{$C/{\rm dof}$} & \colhead{A-D\tablenotemark{b}}
	}
	\startdata
	\swift\ 1\tablenotemark{c} & 2011 Jul 21--Aug 6 & $1.4\pm0.3$ & $6.6\pm1.7$ & 73.5/93 & 39.6\% & $1.5^{+0.5}_{-0.4}$ & - & $5.4\pm1.5$ & 78.8/93 & 90.9\% \\
	{\it XMM} & 2017 Jan 18 & $1.8^{+0.2}_{-0.1}$ & $0.35\pm0.05$ & 200.3/217 & 12.3\% & $0.8^{+0.2}_{-0.1}$ & - & $0.23\pm0.03$ & 215.0/217 & 91.0\% \\
	  & & & & & & $>2.2$ & $0.53\pm0.02$ & $0.34\pm0.11$ & 200.0/216 & 12.8\% \\
	\swift\ 2\tablenotemark{d} & 2017 Nov 21--25 & $1.4\pm0.4$ & $6.3\pm2.5$ & 38.8/40 & 54.2\% & $1.3^{+1.3}_{-0.5}$ & - & $4.7\pm2.1$ & 38.4/40 & 73.1\% \\	 
	\chandra & 2017 Dec 1 & $1.6\pm0.1$ & $7.1\pm0.7$ & 161.2/223 & 23.7\% & $1.5\pm0.2$ & - & $5.6\pm0.7$ & 178.1/223 & 99.9\% \\
	  & & & & & & $>2.5$ & $0.56^{+0.04}_{-0.02}$ & $6.9\pm1.0$ & 160.9/222 & 21.5\% \\
	\enddata
\tablenotetext{a}{The fitted model flux in the energy range 0.3--10\,keV, in units of 10$^{-13}$\,erg\,cm$^{-2}$\,s$^{-1}$.}
\tablenotetext{b}{The percentage result of a goodness-of-fit test using Anderson-Darling statistics.}
\tablenotetext{c}{The combined \swift\ observations 00045596001--00045596006 (first outburst).}
\tablenotetext{d}{The combined \swift\ observations 00045596017 and 000455960018 (second outburst).}
\end{deluxetable*}

\subsection{First brightening}

After a non-detection by \chandra\ in 2005, the first detection of ULX-3 was during a series of observations of NGC~925 with \swift, in six of which the source is detected at fluxes 5--$8\times10^{-13}$\,erg\,cm$^{-2}$\,s$^{-1}$ ($L_{\rm X} = 5.5$--$8.7\times10^{39}$\,erg\,s$^{-1}$ at 9.56\,Mpc). About a month later, two further \swift\ observations failed to detect the source, with the latter establishing an upper limit to the flux of $1.9\times10^{-13}$\,erg\,cm$^{-2}$\,s$^{-1}$. 

Since the first six \swift\ observations (Obs. ID: 00045596001--00045596006) are consistent in flux, we combined the six spectra using {\tt addascaspec} and fitted the stacked spectrum with both an absorbed power-law model ({\tt tbabs*powerlaw} in XSPEC) and a multi-colour disk blackbody model ({\tt tbabs*diskbb}). Of the two models, a hard power-law is preferred, with $\Gamma=1.4\pm0.3$, though we cannot formally reject a disk model. 

\subsection{Between bright epochs} \label{sec:xmm}

Between the first re-brightening and the second, NGC~925 was observed a number of additional times with \swift, and once with \xmm. The single \swift\ detection (Obs. ID: 00045596012) has flux $(7\pm3)\times10^{-14}$\,erg\,cm$^{-2}$\,s$^{-1}$, around an order of magnitude lower than its bright flux, with the remaining upper limits consistent with this value. The \xmm\ detection is at a lower flux of $(3.5\pm0.5)\times10^{-14}$\,erg\,cm$^{-2}$\,s$^{-1}$ ($L_{\rm X} = (3.8\pm0.5)\times10^{38}$\,erg\,s$^{-1}$), and we note that the upper limit placed by the first \chandra\ observation in 2005 is lower still, at $2.8\times10^{-14}$\,erg\,cm$^{-2}$\,s$^{-1}$.

\begin{figure}
	\vspace{5mm}
	\begin{center}
	\includegraphics[width=8cm]{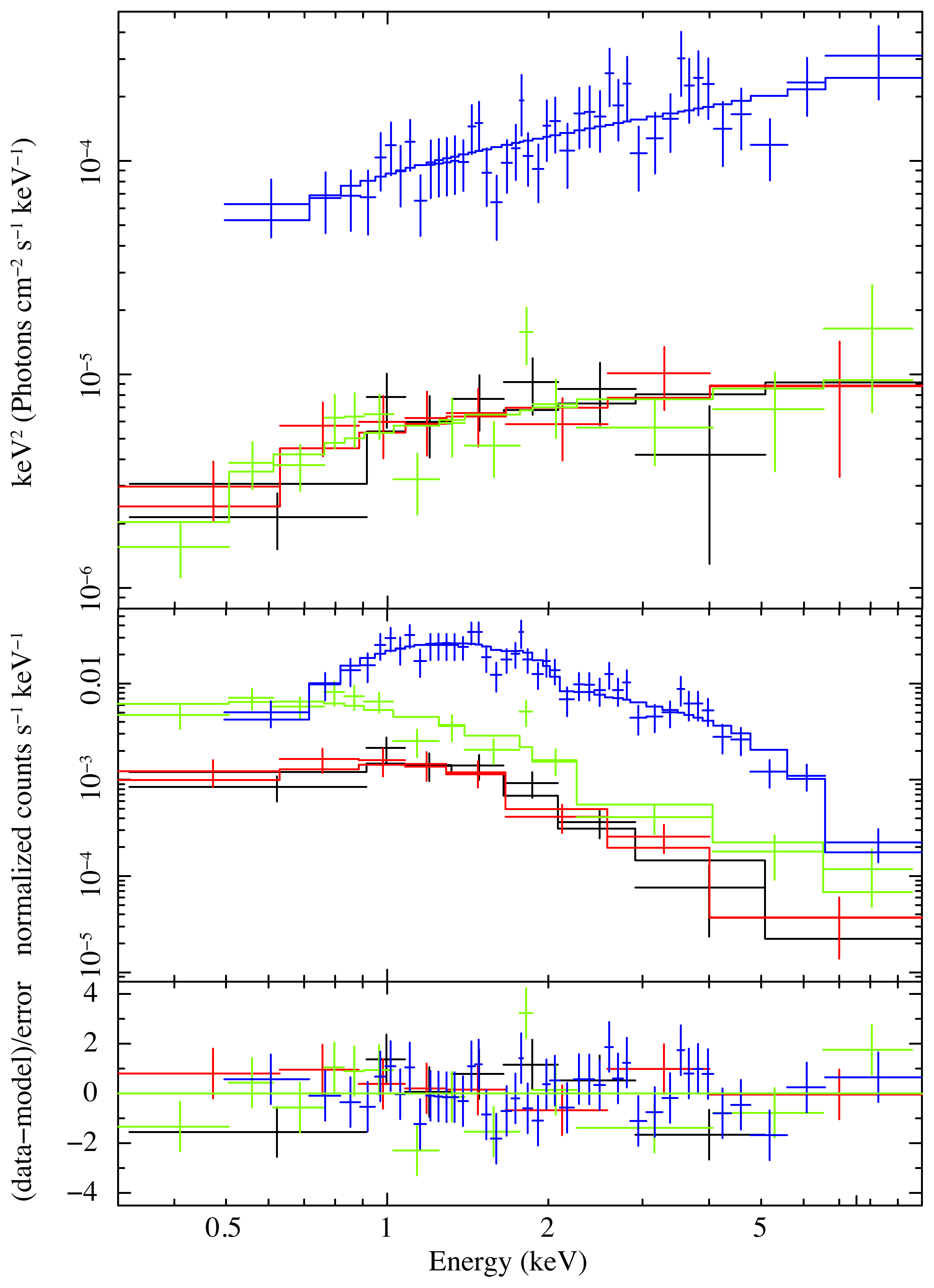}
	\end{center}
	\vspace{-2mm}
	\caption{The spectrum of ULX-3 as observed by \xmm\ in January 2017 (black, red and green for EPIC-MOS1, MOS2 and pn, respectively; approximately 450 net counts), and \chandra\ in December 2017 (blue; approximately 470 net counts), plotted with the best-fitting absorbed power-law for each case (for parameters see Table~\ref{tab:specfitresults}) and binned up for clarity. The top panel shows spectra unfolded through the power-law model, the middle panel shows the counts spectra, and the bottom panel shows the residuals. The combined \swift\ spectra for both the first and second bright states are very similar to the \chandra\ spectrum. \label{fig:spectra}}
\end{figure}

We find the \xmm\ spectrum to be softer than in the high-flux observations, well-fitted with a power-law with $\Gamma = 1.8^{+0.2}_{-0.1}$. A simple disk blackbody model is less favoured, though a broadened disk model ({\tt tbabs*diskpbb}) could fit the data as well as a power-law. However, the data are insufficient to place good constraints on the temperature. We show the spectrum in Fig.~\ref{fig:spectra}, where we compare it with the later \chandra\ detection. 

We find an upper limit on the \nustar\ count rate of $2.46\times10^{-3}$\,ct\,s$^{-1}$ which, for the best-fitting power-law model to the \xmm\ data, gives an upper limit on the 10--20\,keV flux of $4.19\times10^{-14}$\,\,erg\,cm$^{-2}$\,s$^{-1}$. We consider this to be a conservative upper limit, since the spectrum may turn over at higher energies, leading to a far lower flux than if it continues as a power-law.

\subsection{Second brightening} \label{sec:chandra}

The detection of ULX-3 by \chandra\ in 2017 is part of a second high-flux epoch, almost a year after the \xmm\ detection.  It was first seen in \swift, in the latter two of four observations, at similar fluxes to the first outburst (5--$6\times10^{-13}$\,erg\,cm$^{-2}$\,s$^{-1}$). The initial non-detections imply an increase in flux of at least a factor of $\sim$2 over the course of $\sim$40 hours. It was then observed by \chandra\ five days later, at a flux of $(7.1\pm0.7)\times10^{-13}$\,erg\,cm$^{-2}$\,s$^{-1}$ ($L_{\rm X} = (7.8\pm0.8)\times10^{39}$\,erg\,s$^{-1}$).

We combined the two \swift\ observations in which there was a detection as for the first outburst. Both this \swift\ spectrum and the \chandra\ spectrum can be well-fitted with a hard power-law ($\Gamma=1.4\pm0.4$ and $1.6\pm0.1$, respectively), similar to the first \swift\ epoch. We can confidently rule out a disk blackbody model, but once again a broadened disk model could fit the data, although we do not have the high-energy coverage to place good constraints on its parameters.

\subsection{ULX-1 and ULX-2}

The other two known ULXs in NGC~925 were examined in detail in a previous study of data up to 2017 January by \citet{pintore18b} and found to have properties typical of the super-Eddington accreting population of ULXs. Therefore we have not performed an in-depth study of these objects, though we did fit their spectra extracted from the latest \chandra\ observation. 

Both spectra can be well-fitted with power-law models, with ULX-1 having $\Gamma = 1.5^{+0.2}_{-0.1}$ and ULX-2 having $\Gamma = 1.8^{+0.4}_{-0.2}$. The spectral slope for ULX-2 is consistent with that found in the previous study, and the slope for ULX-1 can be considered so when accounting for the inability to detect a high-energy turnover in the \chandra-only data that would soften a power-law model fit. In addition, the model fluxes for both sources are comparable to those found in \citet{pintore18b} when adjusted for the energy range considered. Therefore these sources appear to be persistent and in a similar accretion state to when they were last studied.

\section{Discussion and Conclusions} \label{sec:disc}

Since NGC~925~ULX-3 is observed to be bright on two separate occasions, we can determine that this source is not a single one-off transient event, but is an object that undergoes repeated increases in flux. These high-flux epochs last for at least tens of days, although since we have not observed both the beginning and end of a single high-flux epoch, we cannot say a great deal about their total duration, nor their duty cycle, given the sparse coverage of observations of NGC~925. We do know, however, that ULX-3 exhibits a large dynamic range in flux on month-to-year timescales, with a factor $\sim$26 between the highest-flux detection and the lowest upper limit. 

The lack of a bright optical or infrared counterpart makes it unlikely that this source is a background AGN. At the distance of NGC~925, the faint potential counterpart has absolute $r$-band magnitude $M_r\sim-8$, which is consistent with a supergiant stellar companion. With an apparent magnitude of $m_r\sim21$ it is considerably brighter than most optical ULX counterparts (e.g. \citealt{gladstone13}). As such, this is a good target for further optical/near-infrared observations. Spectroscopic identification of stellar absorption features would allow for the determination of the spectral type of the companion, which has only been possible for a handful of ULX donor stars \citep{motch11,heida15,heida16,heida19}, and make this source an excellent target for attempting radial velocity measurements and thus placing constraints on the mass of the compact object in the system (e.g. \citealt{motch14}).

The limited energy band coverage of \chandra, and the low flux of ULX-3 when observed with \xmm\ and \nustar\ (resulting in a non-detection by the latter), mean that we have so far been unable to find evidence of the spectral turnover at high energies indicative of super-Eddington accretion. Nevertheless, its potential long-term spectral variability, exhibiting a slightly harder spectrum in the 0.3--10\,keV energy range at its highest luminosities and a softer spectrum at a lower luminosity, appears to be more consistent with that seen in some super-Eddington ULXs (e.g. \citealt{pintore12,shidatsu17,walton17}), rather than that expected from the sub-Eddington accretion states of an intermediate-mass black hole (e.g. \citealt{servillat11}). We do, however, note that for existing data the spectral slope measurements for ULX-3 are within error of each other, so further deeper observations are required to confirm this potential spectral variation.

There is not sufficient data for attempting to perform short-term timing analysis such as searching for pulsations, which would be the definitive evidence of the presence of a neutron star in this system. However, the detections of ULX-3 so far available appear to indicate an approximately bimodal flux distribution, as expected for an accreting neutron star that, on occasion, enters a propeller regime with an associated drop in flux. Pulsars with spin periods $\sim$1\,s, as seen in most of the neutron star ULXs discovered to date, are expected to show ratios $\gtrsim$100 between the high and low limiting fluxes (e.g. \citealt{campana01}). However, leakage of a small percentage of the accreting material into the magnetosphere can still occur in a low-flux state, which can reduce the flux ratio to factors of tens (e.g. \citealt{doroshenko11,tsygankov16a}), more consistent with the flux ratio we observe for ULX-3.

Without systematic, high-cadence monitoring of a varying ULX, however, what initially appears to be evidence for the onset of the propeller regime may instead turn out to be a superorbital modulation, which can also result in flux variation of factors of tens in some cases. For example, the ULX M82 X-2 shows dramatic variability that was initially suggested to be due to the propeller effect \citep{tsygankov16a}, but later analysis revealed that this variability was in fact due to a superorbital period \citep{brightman19}. Therefore it is possible that the bimodal flux distribution that we see in ULX-3 is due to a poorly sampled superorbital modulation. 

The existing data is far too sparse for methods such as the Lomb-Scargle periodogram to detect a long-term periodic signal. However, should ULX-3's variability result from a superorbital flux modulation, we can determine from the duration of the high-flux states we have been able to observe that its period is likely to be $\geq$40 days. Simple epoch folding of the existing data gives several potential periods between 70 and 150 days that could be consistent with the existing data (i.e. high and low observed fluxes occur at different points in the phaseogram). Monitoring this source more closely is necessary to better sample the light curve and search for long-term periodicity or quasi-periodicity.

Superorbital periods in ULXs do not tend to show spectral change over the course of a phase cycle -- M82 X-2 is mostly consistent with having a hard spectrum in both high and low flux states \citep{brightman16,brightman19}, and while M51 ULX-7 shows varying hardness over the course of its super-orbital modulation (Brightman et al. in prep.), the source has consistently hard spectra at all fluxes when contamination by soft extended emission at lower fluxes is accounted for \citep{earnshaw16}. In these cases, this implies that the superorbital modulation is not caused by periodic occultation of the neutron star by a warped accretion disk, as we would expect large spectral variation from such a scenario.

On the other hand, study of the more accessible low states of Galactic sources that undergo the propeller effect shows a change from a bright, hard accretion state to a soft, thermal quiescent state, attributed to black-body emission from the surface of the neutron star or its hot spots (e.g. \citealt{reig14,tsygankov16b,fuerst17b}). While ULX-3 does appear to exhibit similar hard-when-bright, soft-when-dim behaviour, the spectrum in the low-flux state is broader, with broadened disc or power-law models preferred over a disc black-body model, and orders of magnitude more luminous than the low states discussed in the aforementioned studies. Therefore we are unlikely to be observing the same propeller effect behaviour in ULX-3.

However, ULX-3's behaviour could potentially be similar to that observed in NGC~5907~ULX-1, whose spectrum is consistently hard in different phases of its super-orbital modulation (though there may be some change in the radial dependence of the disk temperature $p$), but has one lower-flux observation with a softer spectrum, well-fitted with a broadened disc model, which \citet{fuerst17} suggest is associated with refilling of the inner accretion disc after time in the propeller regime. This may indicate the involvement of the propeller effect in ULX-3's light curve, but the lower flux of the \xmm\ observation may still be connected to a superorbital period. 

It is clear that further, denser monitoring of ULX-3 is required to establish the physical cause of its long-term variability. As of August 2019, we have begun an observing campaign with \swift\ in order to improve our coverage of ULX-3's light curve, which we will describe in a future paper along with further planned follow-up. A better understanding of its long-term behaviour will also allow for the effective scheduling of further deep observations to constrain its broadband spectrum in the high-flux state and to better characterise the softer spectrum in the low-flux state.\\

\acknowledgments

We thank our anonymous referee for their useful comments on this paper. This work was supported under NASA contract NNG08FD60C. DJW and MJM acknowledge support from STFC in the form of Ernest Rutherford Fellowships. TPR was funded as part of the STFC consolidated grant ST/K000861/1, and RS was funded by STFC studentship grant ST/N50404X/1. The scientific results reported in this article are based on observations made by the {\it Chandra X-ray Observatory}, as well as archival observations by \xmm, an ESA science mission with instruments and contributions directly funded by ESA Member States and NASA, archival observations by the \nustar\ mission, a project led by the California Institute of Technology, managed by JPL, and funded by NASA, and observations from the \swift\ data archive. 

\vspace{5mm}
\facilities{XMM, CXO, Swift(XRT), NuSTAR}

\software{astropy \citep{astropy13}, CIAO \citep{fruscione06}, FTOOLS \citep{heasarc14}, {\it XMM-Newton} SAS
          }

\bibliography{ngc925ulxpaper}
\bibliographystyle{../aasjournal}

\end{document}